
%
%
%
%
\input harvmac
\nopagenumbers\abstractfont\hsize=\hstitle\rightline{\vbox{\hbox{IMSc
--91/40}}}
\vskip 1in \centerline{\titlefont States of non-zero ghost number in $c<1$
matter}
\bigskip
\centerline{\titlefont coupled to 2-D Gravity}\abstractfont\vskip .5in
\pageno=0
\centerline{Suresh Govindarajan, T. Jayaraman, Varghese John and Parthasarathi
Majumdar\foot{email:suresh, jayaram, john, partha@imsc.ernet.in}}\smallskip
\centerline{The Institute of Mathematical Sciences}
\centerline{C.I.T. Campus, Taramani}
\centerline{Madras 600113}
\centerline{INDIA}
\bigskip
\bigskip

 We study $c<1$ matter coupled to gravity in the Coulomb gas formalism
using the double cohomology of the string BRST and Felder BRST charges. We
find that states outside the primary conformal grid are related to the
states of non-zero ghost number by means
of descent equations given by the double cohomology. Some aspects of the
Virasoro structure of the Liouville Fock space are studied.
As a consequence,
states of non-zero ghost number are easily constructed by ``solving'' these
descent equations.
This enables us to map ghost number conserving
correlation functions involving non-zero ghost number states into
those involving states outside the primary conformal grid.

\vfill
\smallskip

\Date{December 91}
\smallskip
\lref\GRO{U. H. Danielsson and D. Gross, \npb{\bf 366}(1991), 3.}
\lref\FKa{P. Di Francesco and D. Kustasov, \npb{\bf 342}(1990), 589.}
\lref\FKb{P. Di Francesco and D. Kutasov, ``World Sheet and Space time
Physics in two dimensional (super) string theory,'' Princeton preprint
PUPT-1276(1991).}
\lref\DOT{V. Dotsenko, ``Three Point Correlation Functions of the Minimal
Conformal Theories coupled to 2D gravity,'' Paris preprint PAR-LPTHE
91-18(1991).}
\lref\GF{G. Felder, \npb317 (1989), 215.}
\lref\LZ{B. Lian and G. Zuckerman, \plb {\bf 254}(1991), 417.}
\lref\IMM{C. Imbimbo, S. Mahapatra and S. Mukhi,``Construction of Physical
States of Non-trivial ghost number in $c<1$ String Theory,'' Tata preprint
TIFR/TH/91-41(1991).}
\lref\GLI{M. Goulian and B. Li, \prl{\bf 66}(1991), 2051.}
\lref\BMP{P. Bouwknegt, J. McCarthy and K. Pilch, ``BRST analysis of physical
states for 2d gravity coupled to $c<1$ matter,'' CERN preprint
CERN-TH.6162/91 (1991).}
\lref\KIT{Y. Kitazawa, \plb {\bf 265}(1991), 262.}
\lref\KM{M. Kato and S. Matsuda in {\it Advanced Studies in Pure Mathematics},
Vol. 16, ed. H. Morikawa(1988), 205.}
\lref\POL{A. M. Polyakov, \mpl{\bf 6}(1991), 635-644.}
\lref\FF{B. Feigin and D. Fuchs, ``Representations of the Virasoro algebra,''
in {\it Seminar on Supermanifolds} No.5, ed. D. Leites(1988), Univ. of
Stockholm Report No. 25.}
\lref\JD{J. Distler, \npb{\bf 342}(1990), 523.}
\lref\DN{J. Distler and P. Nelson, ``New discrete states of strings near a
black hole,'' Penn preprint UPR-0462T(1991).}
\lref\FUT{S. Govindarajan, T. Jayaraman, V. John, and P. Majumdar, work in
progress.}
\lref\ESG{E. S. Gardner, unpublished.}
\lref\KAL{S. Kalyana Rama, ``New special operators in W-gravity theories,''
Tata preprint TIFR/TH/91-41(1991).}
\lref\SEI{N. Seiberg,``Notes on Quantum Liouville Theory and Quantum Gravity,
,'' Prog. of Theo. Phys., {\bf 102}(1990), 319.}
\lref\POLC{J.Polchinski, ``Remarks on the Liouville field theory,'' Texas
preprint UTTG-19-90, in Proceedings of Strings '90.}
\lref\BER{M.Bershadsky and I.Klebanov, \prl{\bf 65}(1990), 3088; \npb{\bf 360
}(1991), 559.}
\lref\KAED{K. Aoki and E. D'Hoker, ``On the \lio\ approach to correlation
functions for 2-D quantum gravity,'' UCLA preprint UCLA/91/TEP/32(1991).}
\def\plb{Phys. Lett. {\bf B}}
\def\prl{Phys. Rev. Lett.}
\def\mpl{Mod. Phys. Lett. {\bf A}}
\def\npb{Nuclear Phys. {\bf B}}
\def\ket#1{| #1 \rangle}
\def\melt#1#2#3{\langle #1 \mid #2 \mid #3 \rangle}
\def\lfr#1#2{\textstyle{#1 \over #2 }}
\def\al{{\alpha}}
\def\pa{{\partial}}
\def\det{\hbox{det$\,$}}
\def\lio{Liouville}
\def\vm#1#2{\ket{v_{{#1},{#2}}}_M}
\def\um#1#2{\ket{u_{{#1},{#2}}}_M}
\def\wm#1#2{\ket{w_{{#1},{#2}}}_M}
\def\vl#1#2{\ket{v_{{#1},{#2}}}_L}
\def\gh{c_1\ket0_{gh}}

\def\dotm#1#2{\vm{#1}{#2}\otimes\vl{-{#1}}{#2}\otimes\gh}

\newsec{Introduction}

 While the matrix model approach to $c\leq1$  matter coupled to gravity
has made considerable progress the same level of clarity and
computability has not yet been achieved  in the continuum Liouville
approach to these theories. The true Liouville theory is certainly
not a free field theory.  However important progress has been made
by recognising that at least in some domain the Liouville theory behaves
like a modified free field theory \POLC\SEI. A significant success of this
approach has been the calculation of the torus partition function of minimal
models coupled to 2-d gravity \BER. Further Polyakov has pointed out that at
least some part of the target space scattering amplitudes may be
obtained by purely free field techniques\POL.

But even in this free field approach major questions still remain to be
answered
satisfactorily. In particular the issue of the identification of
physical operators and the evaluation of their correlation functions
is still not completely resolved.
Two important steps have however been taken in addressing these  questions.
Lian and Zuckerman \LZ\ have shown that there exist an infinite number of
BRST invariant states in $c<1$ theories coupled to gravity. The
Liouville momenta of these states are such as to provide the
gravitational dressing of the matter null states of the minimal model.
Further there is one state at ghost number $\pm n$ for every matter Virasoro
representation whose Liouville
momenta are  $\beta > \beta_0$ for ghost number $+n$ states and $\beta <
 \beta_0 $ for ghost number $-n$ states, where $\beta_0$ is the
background charge of the Liouville theory.
Using the theorem of Lian and Zuckerman, an explicit construction of
the ghost number $\pm 1$ states has been given by Imbimbo, Mahapatra and Mukhi
\IMM.
The three-point correlation functions of the
states of ghost number zero have been computed using the Coulomb gas
representation of the matter part of the theory by a number of authors
\GLI\DOT\KIT\KAED. Surprisingly,in these calculations, free field techniques
supplemented by the technique of negative number of screening contours in the
\lio\ sector are sufficient to produce results in agreement with those
 computed from the matrix models.
  Dotsenko and Kitazawa \DOT\KIT\ noticed that for the $c<1$ minimal models
coupled to gravity the
matter vertex operator momenta could be continued to values outside the
primary grid, with non-zero results for the  correlation function. These
operators have the same Liouville momenta as the Lian-Zuckerman
states,at least in part, for the appropriate choice of gravitational dressing.
Thus there appears to be two types of states,which we refer to as the
LZ (Lian-Zuckerman)and DK (Dotsenko-Kitazawa) type states. The corresponding
operators are referred to as LZ and DK operators.
The ghost number zero operators
are of course the same in both LZ and DK form.

In this letter we study the $c<1$ models coupled to Liouville
using the Coulomb Gas formalism for the matter part of the theory.
We show using the rederivation of the Lian-Zuckerman
theorem by Bouwknegt,McCarthy and Pilch \BMP\ that the DK type states are
related to LZ states by a set of descent equations in a doubly graded
complex  associated  with two BRST operators. One of these is the usual string
BRST operator for the theory and  the other is the BRST operator
associated with the screening operators of the Coulomb gas formalism
as in the work of Felder\GF. For this construction, we also need a more
detailed knowledge of the Virasoro structure of  the \lio\ Fock space.
This is studied using the techniques of Kac determinants for Fock spaces
ala Kato and Matsuda \KM.
 This in turn provides us with an algorithm
to explicitly construct the LZ type states,beginning with DK type states.
The DK type states are, of course, trivial to construct. We illustrate our
construction of LZ states from DK states. We then demonstrate the use of these
descent
equations to show  that the correlation function of LZ operators may be
reduced to correlation of DK operators.
\newsec{Matter and Liouville Fock Spaces}
Following \BMP, we consider two scalars $\phi^M$ and $\phi^L$ with charges
$\al_0$
and $\beta_0$ respectively at infinity. The corresponding energy-momentum
tensors
are given by
\eqn\estress{\eqalign{
T^M &= -\lfr14\pa\phi^M\pa\phi^M + i \al_0 \pa^2\phi^M\quad,\cr
T^L &= -\lfr14\pa\phi^L\pa\phi^L + i \beta_0 \pa^2\phi^L\quad,
}}
with central charges $c_M=1-24\al_0^2$ and $c_L=1-24\beta_0^2$. For the
$(p,p+1)$ unitary minimal models
$$
\al_0^2=\lfr1{4p(p+1)}~ {\rm and}~\beta_0^2=
-\lfr{(2p+1)^2}{4p(p+1)}\quad.
$$
 The vertex operators $e^{i\al\phi^M}$ and
$e^{i\beta\phi^L}$ have conformal weights $\al(\al-2\al_0)$ and
$\beta(\beta-2\beta_0)$ respectively. The usual screening charges for matter
are
$$
\al_+=\sqrt{\lfr{p+1}p}~{\rm and}~\al_-=-\sqrt{\lfr p{p+1}}\quad.
$$
The screening charges for the \lio\ sector are given by
$$
\beta_+=i\al_+~{\rm and}~\beta_-=-i\al_-\quad.
$$
Following Felder\GF, we consider the complex of Fock spaces
(hereafter referred to as the Felder complex)
in the matter sector given by $\bigoplus ~F_{m',\pm m+2np}$, where $F_{m',m}$
is the Fock space built over the primary associated with the vertex operator
$e^{i\al_{m',m}\phi}$. Here
$$
\al_{m',m}=\lfr{1-m'}2\al_- + \lfr{1-m}2\al_+\quad.
$$
We will also need the dual Fock space obtained by $F_{-m',-m}$.
There exists an identity under the change of label given by $(m',m)\rightarrow
(m'+p+1, m+p)$ for the matter sector and $(m',m)\rightarrow(m'+p+1,m-p)$ in the
\lio\ sector. These two seemingly distinct labels refer to the same vertex
operator. This identity will prove to be useful later.
There is one such Felder complex for every $m',m$ restricted to the range
$1\leq m'\leq p$ and $1\leq m \leq (p-1)$. The screening operators
$Q_m^+ = \int \prod_{i=1}^m dz_i e^{i\al_+\phi(z_i)}$ and similarly $Q_{p-m}^+$
act on these Fock spaces. The irreducible Virasoro module $L(m',m)$ (for a
given $c_M$ labelled by $p$) is given by $Ker~Q_m^+/Im~Q_{p-m}^+$ on this
complex. We shall refer to the screening operators loosely as $Q_F$ except
when necessary.
 We also have the Fock spaces of the \lio\ and ghost sectors denoted
by $\CF(\beta)$ and $\CF(gh)$ respectively. The string BRST operator $Q_B$
given by
\eqn\ebrst{
Q_B = \oint :c(z) (T^M(z) + T^L(z) + \lfr12 T_{gh}(z)):
}
acts on the tensor product $\CF(\al)\otimes\CF(\beta)\otimes\CF(gh)$.

We now proceed to consider the Virasoro
structure of these Fock spaces. In the matter case, this is encoded in
the diagram given in Figure 1. In each Fock tower, $v_{m',m}$ is
the highest weight
vector associated with the operator $e^{i\al_{m',m}}$, with an irreducible
Virasoro module under it. $u_{m',m}$ is a level $(p+1-m')(p-m)$
 singular vector over $v_{m',m}$ with an irreducible Virasoro module
over it.
 It is generically given by an expression of the form
\eqn\enullu{
{\CL}_{(p+1-m')(p-m)}\vm{m'}m = ( L_{-(p+1-m')(p-m)} + \cdots )\vm{m'}m\quad,
}
where the ellipsis refer to a polynomial of $L_{-n}$'s at level
$(p+1-m')(p-m)$.
The second Virasoro null vector at level $m'm$ is identically vanishing in the
Fock space. At that level, we have a Fock space state, $w_{m',m}$ of level
$m'm$ which cannot be obtained by the action of $L_{-n}$'s on $v_{m',m}$.
$w_{m',m}$ can be obtained as the limit(see for instance \GRO)
\eqn\enullw{
\ket{w_{m',m}} = \lim_{(\al\rightarrow \al_{m',m})} \lfr1{(\al - \al_{m',m})}
\CL_{-m'm} \ket{v^\al}\quad,
}
where $v^\al=e^{i\al\phi^M}$. Thus, $w_{m',m}$ is the analog of the discrete
primary in the $c_M=1$ case. There
these states became BRST closed on being given an appropriate
\lio\ dressing. However, we will find that they play a slightly different
role here.The pattern of $v_{m',m}$ with a null vector of  type $u_{m',m}$ and
a Fock state $w_{m',m}$ is repeated in every Fock tower.\foot{The $v,~u,~w$
in our notation correspond to the $v_0,~u_1,~w_0$ in Felder's notation.}
For details, see \GF.

We would like to understand the null state structure of the \lio\ primaries.
For that, we note that Kato and Matsuda\KM\ have provided the connection
between Virasoro secondaries in the Fock space and the Fock states(those
created by the action of creation and annihilation operators). We may
write
\eqn\ekato{
\CL_{I} \ket{\al_{m',m}} = \sum C_{IJ}a_{J}\ket{\al_{m',m}}\quad,
}
where $\CL_{I}$ are products of Virasoro raising operators of level $N$, $I=1,
\ldots,p(N)$\foot{$p(N)$ refers to the number of partitions of $N$.}.
$a_J$ are the products of Fock creation operators of level $N$
with $J=1,\ldots,p(N)$. $C_{IJ}$ is a matrix representation of the Virasoro
secondaries in term of the Fock states.
The zeros of the \det$C_{IJ}$ corresponds to the null vectors of the Virasoro
structure which do not map into the Fock space. The \det is given by the
expression
\eqn\edeta{
   det\,C_{IJ} = {\rm const.}{\prod_{1\leq m'm\leq N}}
 (\al -\al_{m',m})^{p(N-m'm)}\quad{\rm for}\quad m',m>0\quad,
}
where the zeros of the determinant are explicitly shown.
The non-vanishing null is obtained by examining the det for the dual Fock
space.
\eqn\edetb{
   det\,\tilde{C}_{IJ} = {\rm const.}{\prod_{1\leq m'm\leq N}}
 (\al -\al_{m',m})^{p(N-m'm)}\quad{\rm for}\quad m',m<0\quad.
}
For example, consider the matter primary given by $\al_{m',m}$. Note that
\edeta\ and \edetb\ have a zero at levels $m'm$ and $(p+1-m')(p-m)$
respectively. Hence, it has a vanishing null at level $m'm$ and a non-vanishing
null at level $(p+1-m')(p-m)$. They interchange their roles for the dual Fock
primary which is given by $\al_{p+1-m', p-m}$. Now, we shall apply this
to the \lio\ Fock primary labelled by $\beta_{m',m}$
We consider three cases.

\noindent
Case 1: $\beta_{-m',m}$  with $0<m'<p,~0<m<(p-1)$

Since \edeta\ and \edetb\ do not have any zeros for these values or the
corresponding dual, there is no null over these primaries. This is true in all
the cases when the \lio\ dresses a matter primary inside the conformal grid.
This includes the case of the cosmological constant\ESG.

\noindent
Case 2: $\beta_{m',m}$ with $0<m'<p,~0<m<(p-1)$

Now \edeta\ has a zero at level $m'm$ indicating a vanishing null. The dual
has a non-vanishing null at the same level. Note that the $m',~m$ labels of the
dual are both negative and cannot be both made simultaneously positive by means
of the symmetry discussed earlier.  For the
case of the vanishing null, $\beta<\beta_0$ and $\beta>\beta_0$ for the other
case. This is the case studied in \IMM.

\noindent
Case 3: $\beta_{m',m}$ with $0<m'<p,~m>(p-1)$

Now \edeta\ has zeros at levels $m'm$ and $(n(p+1)+m')(m-np)$ where
$n$ is fixed so that $(m-np)$ lies inside the conformal grid.
Thus, in this case
we have two vanishing nulls. The dual has two non-vanishing nulls given by
the zeros of \edetb. They are at the same levels as in the vanishing case.
Note again that the $m',~m$ of the dual are both negative and can never be
made simultaneously positive. Once again, for the
case of the vanishing null, $\beta<\beta_0$ and $\beta>\beta_0$ for the other
case.

It appears possible to extend this argument to completely determine the
Virasoro structure of the \lio\ sector. It would appear that \lio\ should
belong to the ${\rm III}_+(-)$ and ${\rm III}_+(+)$ case of Feigin and
Fuchs\FF.
 We shall however
leave detailed considerations for the future\FUT. For the work at hand, the
first two nulls over the \lio\ Fock primary are sufficient.
\newsec{Double Cohomology of $Q_B$ and $Q_F$}
The cohomology of the string BRST for $c<1$ matter coupled to \lio\ is the
object of interest here. The tensor product of the Fock spaces $\CF(\al)\otimes
\CF(\beta)\otimes\CF(gh)$ is labelled by two gradings. One is the usual
ghost number $G$($b$ has ghost number $-1$, $c$ has ghost number $+1$ and
the state $\gh$ is chosen to have ghost number $0$). The other grading related
to the Felder cohomology is given by the distance of the given Fock tower from
the central tower in the Felder complex. We shall call it tower number, with
positive to the right of the central tower and negative to the left(refer
Figure 1). One may study this cohomology directly on the relevant Fock
spaces before truncating the matter Fock space to the Felder cohomology. There
is one obvious class of states that are in the $Q_B$ cohomology. They are
given by the states
\eqn\estatea{
\ket{\al_{m',m\pm2np}}_M\otimes\ket{\beta}_L\otimes\gh\quad,
}
where $\beta$ is chosen to obey the gravitational dressing condition
$$
\al(\al-2\al_0) + \beta(\beta-2\beta_0) =1\quad.
$$
So all the $v_{m',m}$'s from every Fock tower in the Felder complex suitably
dressed are in the cohomology. Under the
action of $Q_F$, the states given in \estatea\ are either exact or non-closed
except in the central Fock tower. However, they map to or map from states
that are not in the $Q_B$ cohomology.  Thus the set of
states in \estatea(which are in fact the DK states)
form representatives of the cohomology classes denoted by
\eqn\ecoh{
H^{(n)}(
H^{(0)}_{rel}(\CF(\al)_M\otimes\CF(\beta)_L\otimes\CF_{gh},~Q_B),~Q_F)\quad.
}

Bouwknegt, McCarthy and Pilch\BMP\ have shown that there is
an isomorphism between
cohomology classes given by
\eqn\ecoha{\eqalign{
H^{(n)}_{rel}(
H^{(0)}(\CF(\al)_M&\otimes\CF(\beta)_L\otimes\CF_{gh},~Q_F),~Q_B)\cr
&\simeq
H^{(n)}(
H^{(0)}_{rel}(\CF(\al)_M\otimes\CF(\beta)_L\otimes\CF_{gh},~Q_B),~Q_F)\quad.
}}
Rewriting the left hand side as
\eqn\ecohb{
H^{(n)}_{rel}(L(m',m)\otimes\CF(\beta)_L\otimes\CF_{gh},~Q_B)\quad,
}
we find that the states belonging to \ecohb\ are the LZ states, the objects
of interest. Here, we take the isomorphism further by explicitly relating
the DK states to the LZ states by means of descent equations in the double
complex of $Q_B$ and $Q_F$. This isomorphism in particular, would relate
states at non-zero ghost number(LZ) to those of zero ghost number(DK). We
illustrate the descent for ghost number $-1$. Take a DK state of tower number
$-1$ given by $\dotm{m'}{2p-m}$.  The choice of \lio\ dressing here belongs
to case 2 of the previous section with a vanishing null. Then
\eqn\edesa{\eqalign{
Q_F\ket{DK}&=Q_F\dotm{m'}{2p-m}\cr
&=\um{m'}m\otimes\vl{-m'}{2p-m}\otimes\gh
}}
Then one can construct a LZ state such that
\eqn\edesb{
Q_B \ket{LZ}=\um{m'}m\otimes\vl{-m'}{2p-m}\otimes\gh\quad.
}
\edesa\ and \edesb\ imply that
\eqn\edesc{
Q_B\ket{LZ}= Q_F\ket{DK}\quad.
}
Note further that
\eqn\edesd{
Q_F\ket{LZ}=0\quad,\quad Q_B\ket{DK}=0\quad.
}
\edesc\ and \edesd\ give the descent equations for ghost number $-1$. This
procedure works for all the examples constructed in \IMM. The general
descent equation for ghost number $-n$ is given by
\eqn\edese{\eqalign{
Q_B\ket{LZ}  &= Q_F\ket{I_1}\quad,\cr
Q_B\ket{I_1} &= -Q_F\ket{I_2}\quad,\cr
             &~\vdots \cr
Q_B\ket{I_{n-1}}&=(-)^{n+1} Q_F\ket{DK}\quad.
}}
Such a descent equation follows from the statement that
$(\ket{LZ}+\ket{I_1}+\ldots+\ket{DK})$ is closed under $(Q_B-(-)^GQ_F)$.
Note that the sum of the tower number and  ghost number is invariant for
$\ket{LZ},~\ket{DK},~\ket{I_m}$.
As we shall see later, for ghost number $+n$, the situation
is different. There, the descent equations are valid but vacuous.

Since the DK states
are easy to construct, one can now ``solve'' \edese\ to obtain the LZ state.
This now provides a systematic procedure for generating  LZ states of negative
ghost number.  Establishing  every equation in \edese\ will require
knowing the appropriate null state equations in the matter and \lio\
Fock towers. We shall demonstrate this
construction for ghost numbers $-1,~-2$ in the next section.
Interestingly, the condition $\beta<\beta_0$ as the dressing
of the DK state is required in ``solving'' \edese, which is also the condition
given by Lian and Zuckerman. This always corresponds to
choosing the vanishing null over the \lio\ primary.

The states of positive ghost number require a slightly different construction.
All the positive ghost number states are obtained from the oscillator states
$\ket w$ which appear in place of the vanishing Virasoro null vector(in
the matter sector). As pointed out earlier, these states can be defined by
means of a limit procedure. Note that this is equivalent to defining them
by means of Fock creation operators.

We shall first describe the situation for ghost number $+1$. We have
\eqn\eposa{\eqalign{
\ket{DK}&= Q_F \wm{m'}m\otimes\ket{\beta}_L\otimes\gh\quad,\cr
\ket{LZ}&= Q_B \wm{m'}m\otimes\ket{\beta}_L\otimes\gh\quad,\cr
}}
where the Liouville dressing is fixed to be $\beta>\beta_0$ in order to agree
with Lian and Zuckerman. Unlike
in the negative ghost number case, we find that this choice is not essential
for our construction to work. The state $\ket{LZ}$ in \eposa\ is not truly
$Q_B$ exact
since $w_{m',m}$ is not in the Felder cohomology.\foot{This is very much like
the zero momentum dilaton in the bosonic string where the dilaton can be
written
as $Q_B(c_0-\bar{c}_0)\ket0$. Yet it is not a truly BRST exact state  since
$(c_0 - \bar{c}_0)\ket0$ is not annihilated by $(b_0-\bar{b}_0)$ and
hence is not an allowed state in the Hilbert space.}
A similar argument holds for $\ket{DK}$ which is not truly $Q_F$ exact.
It is easy to see that state with $w_{m',m}$ in the matter sector would become
a discrete primary in the $c=1$ limit and would be $Q_B$ closed.

This procedure is easy to generalise for ghost number $+n$.
\eqn\eposb{\eqalign{
\ket{DK} &= Q_F \ket{W^{(1)}}\quad,\cr
\ket{I_1} &= Q_B \ket{W^{(1)}}\quad,\cr
\ket{I_1} &= Q_F \ket{W^{(2)}}\quad,\cr
\ket{I_2} &= Q_B \ket{W^{(2)}}\quad,\cr
         &=~~\vdots\quad,\cr
\ket{I_{n-1}} &= Q_F \ket{W^{(n-1)}}\quad,\cr
\ket{LZ} &= Q_B \ket{W^{(n-1)}}\quad,\cr
}}

where  $\ket{W^{(m)}}= \CL^c\ket{w^{(m)}}\otimes\ket\beta_L\otimes\gh$ and
$w^{(m)}$ is the first oscillator state associated with
matter Fock tower $m$ towers to the left of $\ket{DK}$(i.e., tower number
$(n-m)$). Further, $\CL^c$ is a polynomial of ghost number $(m-1)$ in
$L_{-n}$'s and
appropriate $c_{n}$'s.

Finally, note that we have also used the isomorphism
between the irreducible Verma modules $L(m',m)$ and $L(p'-m',p-m)$.
 In the construction of the LZ states
we have used  the first Virasoro irrep for one half of the states
and the other irrep for the rest.
\newsec{Examples}
	In this section, we present explicit construction of LZ
states from DK states. We obtain examples for states of ghost number
$\pm 1$ and $\pm 2$ to illustrate the construction. We compare the ghost number
$\pm 1$ states obtained by our construction to those in \IMM.

First, we construct the ghost number $-1$ state for $(m',m)=(-2,-1)$ for
arbitrary $c_M$. The descent equation is given by
\eqn\edesca{
    Q_B \ket{LZ} = Q_F \ket{DK}
}
The DK state is given by
\eqn\eega{
\ket{DK}
= \vm{-2}1 \otimes \vl21 \otimes \gh \quad.
}
We then have
\eqn\eegb{\eqalign{
Q_F\ket{DK} &= \um{-2}{-1}\otimes\vl21\otimes c_1\ket0_{gh}\quad,\cr
         &= (L^M_{-2} -t L_{-1}^{M~2})\vm{-2}{-1}\otimes\vl21\otimes
c_1\ket0_{gh}\quad,
}}
where $t=\lfr p{p+1}$.
The corresponding LZ-state is given by solving the descent equation \edesca\
\eqn\elza{
\ket{LZ} = ( t(L^M_{-1}-L^L_{-1})b_{-1}+ b_{-2} )
\vm{-2}{-1}\otimes\vl21\otimes c_1\ket0_{gh}\quad.
}
In checking equation \edesca, one encounters the Liouville null state
 given by
\eqn\enulla{
(L^L_{-2} + t L^{L~2}_{-1})\vl{-2}3 \quad,
}
which is  zero since the choice of Liouville dressing is made so as to have a
vanishing null. This also corresponds to Lian and Zuckerman's choice of
$\beta<\beta_0$.

We shall now construct a ghost number $-2$ LZ state. For convenience, we choose
the case of $c_M=0$. The descent equation is
\eqn\edescb{\eqalign{
    Q_B \ket{LZ}& = Q_F \ket{I}\quad,\cr
    Q_B \ket{I} &= -Q_F \ket{DK}\quad.
}}
Figure 2 shows the location of the matter part of these states in the
Felder complex.
Consider the DK state obtained from dressing $\vm25$. ``Solving'' \edescb\
and after tedious but straightforward algebra, we obtain
\eqn\elzb{\eqalign{ \ket{DK} &=\vm25\otimes\vl{-2}5\otimes\gh\quad,\cr
 \ket I &= \CL^b_{4} \vm23\otimes\vl{-2}5\otimes\gh\quad,\cr
 \ket{LZ} &= \CL^{2b}_{5}\vm21\otimes\vl{-2}5\otimes\gh\quad,
}}
where
\eqn\eec{\eqalign{
\CL^b_{4} &= \lfr{94}3 b_{-4} + b_{-3}(\lfr{61}3 L^L_{-1} +3L^M_{-1})\cr
           &+ b_{-2}(4L^L_{-2}- 4L^M_{-2} +\lfr{20}3L^{L~2}_{-1})\cr
           &+b_{-1}(-3L^L_{-3} -\lfr{41}3L^M_{-3}-\lfr{20}3L^L_{-1}L^M_{-2}\cr
   &+\lfr{20}3L^M_{-2}L^M_{-1} + L^{L~3}_{-1} -L^{L~2}_{-1}L^M_{-1}
   + L^{L}_{-1}L_{-1}^{M~2}-L^{M~3}_{-1})\quad,
}}
and
\eqn\eed{\eqalign{
\CL^{2b}_{5} &= -\lfr43b_{-4}b_{-1} + 4b_{-3}b_{-2}
+b_{-3}b_{-1}(\lfr23L^L_{-1}-15L^M_{-1})\cr
              &+ b_{-2}b_{-1}(-4 L^L_{-2} -\lfr23L^{L~2}_{-1}
-6L^L_{-1}L^M_{-1} + L^{M~2}_{-1})\quad.
}}
These expressions are unique upto $Q_B$ exact pieces that are also $Q_F$
closed. In the process of checking that \elzb\ satisfies \edescb,
vanishing nulls encountered in both the \lio\ and matter sectors.
The relevant null state equations are
\eqn\enullb{\eqalign{
\um23 &= \CL_4^M \vm23\quad,\cr
(L^M_{-2} -\lfr32 L_{-1}^{M~2})\vm21 &=0\quad,\cr
 \CL_4^L\vl{-2}5=0\quad,& \CL_3^L\vl{-2}5=0\quad,
}}
where\foot{The Liouville state $\vl{-2}5\equiv\vl13$ has two vanishing nulls
over it as
given by Case 3 in the section 2.}
$$\eqalign{
\CL^M_{4}&=4L^M_{-4}-4L^M_{-3}L^M_{-1} -4L^{M~2}_{-2} +
\lfr{20}3L^M_{-2}L^{M~2}_{-1} - L^{M~4}_{-1}\quad,\cr
\CL^L_{4}&=\lfr{76}3L^L_{-4}+\lfr{52}3L^L_{-3}L^L_{-1} +4L^{L~2}_{-2} +
\lfr{20}3L^L_{-2}L^{L~2}_{-1} + L^{L~4}_{-1}\quad,\cr
\CL^L_3 &= L_{-3}^L +\lfr12 L_{-2}^L L_{-1}^L +\lfr1{12}L_{-1}^{L~3}\quad.
}$$
This procedure can be used to generate LZ-states at arbitrary negative ghost
number. Of course, in practice this entails the use of higher null state
equations that rapidly increase in complexity which makes computations tedious.

As mentioned earlier, a slightly different procedure is required to
construct states at positive ghost number. We shall now illustrate the
construction for ghost numbers $+1,~+2$. Consider the ghost number $+1$ case.
We begin by considering the $w_{1,1}$ in the matter $(1,1)$ Fock tower.
\eqn\ewa{
\wm11=\lim_{\al\rightarrow \al_{1,1}} \lfr{1}{\al -\al_{1,1}}
L^M_{-1}\vm11\quad.
}
The LZ and DK states are obtained from the above matter state(appropriately
dressed by a Liouville primary). They are
\eqn\elzc{\eqalign{
\ket{LZ}_{+1} &= Q_B \wm11\otimes\vl{-1}{-1}\otimes\gh\cr
              &= -4 \al_0 c_{-1}\vm11\otimes\vl{-1}{-1}\otimes\gh\quad,\cr
\ket{D} &=\vm1{-1}\otimes\vl{-1}{-1}\otimes\gh\quad.
}}
This agrees with the corresponding LZ-state given in \IMM.
We now construct another LZ-state at ghost number $+1$. Consider the $w_{2,1}$
in the Fock tower  $(m',m)=(2,1)$.
\eqn\ewb{
\wm21=\lim_{\al\rightarrow \al_{2,1}} \lfr{1}{\al -\al_{2,1}}
( L^M_{-2}-\lfr3{2(2\Delta+1)}L^{M~2}_{-1}) \vm21\quad.
}
where $\Delta$ is the weight of $\vm12$.
The corresponding LZ and DK states are given by
\eqn\elzd{\eqalign{
\ket{LZ}_{+1} &\propto c_{-2}\vm21\otimes\vl{-2}{-1}\otimes\gh\quad,\cr
\ket{DK} &=\vm2{-1}\otimes\vl{-2}{-1}\otimes\gh\quad.\cr
}}
This differs from the corresponding state in \IMM\ by a $Q_B$-exact piece. The
exact part is obtained by acting with $Q_B$ on the level two Virasoro secondary
(orthogonal to the non-vanishing Virasoro null) over the primary $\vl{-2}{-1}$.
Note that since this term is given by $Q_B$ acting on a state in the Felder
cohomology, it is truly exact.

We now will construct a ghost number $+2$ state. We shall do this for the
case of $c_M=0$ and $(m',m)=(1,1)$. Figure 2 shows the location of the matter
part of these states in the Felder complex.
Consider the DK state
\eqn\edk{
\ket{DK}= \vm1{-3}\otimes\vl{-1}{-3}\otimes\gh\quad.
}
Following our construction, $Q_F\wm1{-1}= \vm1{-3}$ with
\eqn\ewc{
\wm1{-1}=\lim_{\al\rightarrow \al_{1,-1}} \lfr{1}{\al -\al_{1,-1}}
 \CL^M_{4}\vm1{-1}\quad,
}
where $\CL_4^M$ is as given after \enullb.
The state $\ket I$ is obtained as follows
\eqn\eia{\eqalign{
\ket{I}_{+1} &=  Q_B \wm1{-1}\otimes\vl{-1}{-3}\otimes\gh\cr
&=\CL^c_{-4}\vm1{-1}\otimes\vl{-1}{-3}\otimes\gh\quad.
}}
where
$$
\CL^c_{4} =-\lfr5{\sqrt6}
\{ c_{-1}(2L^M_{-3} -\lfr{20}3L^M_{-2}L^M_{-1}+2L^{M~3}_{-1} )
+ c_{-2}(-2L^M_{-2} + \lfr73L^{M~2}_{-1}) -\lfr{10}3c_{-3}L^M_{-1} \}
$$

The intermediate oscillator state(again replacing a vanishing null) encountered
is given by using
\eqn\ewd{
\ket{I}_{+1}=  Q_F \ket{\tilde w_{1,1}}\otimes\vl{-1}{-3}\otimes\gh\quad.
}
We obtain
\eqn\ewe{\eqalign{
\ket{\tilde w_{1,1}} &= \CL^c_{-4}\wm11\cr
&= \lim_{\al \rightarrow\al_{1,1}} \lfr1{\al-\al_{1,1}} \CL^c_{-4}
L^M_{-1}\vm11\quad.
}}
The LZ state is given by
\eqn\elze{\eqalign{
\ket{LZ}_{+2} &= Q_B \ket{\tilde w_{1,-1}}\cr
=\lfr56 \{28c_{-4}&c_{-1} + 56 c_{-3}c_{-2} -36c_{-2}c_{-1}L^M_{-2}\}
\vm11\otimes\vl{-1}{-1}\otimes\gh.
}}

It is straightforward to check that the norm
$\phantom{}_{-2}\melt{LZ}{c_0}{LZ}_{+2}$ is non-zero. This is a simple check to
show that the states are in the
cohomology.

\newsec{Correlation Functions}
\def\corr#1{\langle {#1}\rangle}
\def\oplz#1#2#3{\Phi_{#1,#2}^{(#3)}}

 The isomorphism of LZ states and DK states can now be used in interpreting  at
least some of the correlation functions calculated by Dotsenko and Kitazawa.
Let us begin
with a simple example of the three-point function of operators with  ghost
number
$+1$ and $-1$ and a ghost number zero operator.We show that this can be
converted
into  a three-point function with all the operators in the DK form.
Denote by
$\Phi_{m',m}^{(n)}$ the operator of ghost number $n$ whose matter part is a
state with charge $\al_{m',m}$ that creates an LZ state from the  vacuum.
We  write the three point function
\eqn\ecothree{
X \equiv \corr{\oplz{m'}{m}{-1} \oplz{n'}n0 \oplz{k'}k{+1}}
}
We can now write the operators explicitly in the Coulomb gas language.
We may write $\oplz{n'}n0$ such that in the matter part we have a suitably
screened vertex operator ala Felder defined as
\eqn\ecofour{
V_{n',n}^{r',r}(z)=\int V_{n',n}(z) V_{\al_-}(u_1) \ldots V_{\al_-}(u_{r'})
V_{\al_+}(v_1)\ldots V_{\al_+}(v_r) \prod_{i=1}^{r'}du_i \prod_{j=1}^r dv_j
\quad,
}
where $V_{n',n}=e^{i\al_{m',m}\phi^M}$, $V_{\al_\pm} =e^{i\al_\pm}$,
$2r'=n'+m'-k'-1$ and $2r=n+m-k-1$.
Writing $Q_B=\oint J_B(z)$ and using the construction of the ghost number $+1$
state given earlier we obtain
\eqn\ecofive{
X = \corr{\oplz{m'}{m}{-1} \oplz{n'}n0 \oint J_B(z) W_{k',k}(0)}\quad,
}
where $W_{k',k}$ is the appropriate  state associated with the Fock secondary.
We can now deform the contour so that it picks up the contribution from the
other vertex operators. There is no contribution from
$\oplz{n'}n0$. The only non-zero contribution arises from the
contour enclosing $\oplz{m'}m{-1}$. Thus, we obtain
\eqn\ecosix{
X = \corr{\oint J_B(z) \oplz{m'}{m}{-1} \oplz{n'}n0  W_{k',k}}\quad.
}
On using the descent equation \edesc, we get
\eqn\ecoa{
X = \corr{Q_{p-m} \oplz{m'}{-m+2p}{D} \oplz{n'}n0 W_{k',k}}\quad,
}
where $\oplz{m'}{-m+2p}{D}$ is the corresponding DK operator. We can now
deform the contour of $Q_{p-m}$ through to the right(picking up phase factors)
to give
\eqn\ecob{\eqalign{
X &= \corr{ \oplz{m'}{-m+2p}{D} \oplz{n'}n0 Q_{p-k}W_{k',k}}\cr
&= \corr{ \oplz{m'}{-m+2p}{D} \oplz{n'}n0 \oplz{k'}{-k}D}\quad.
}}
The screened vertex operator associated with the matter part of $\oplz{n'}n0$
is actually modified to $V_{n',n}^{r',r}$ where now $2r = n-m+k-1$ and $r'$
remains unchanged.
Thus we have converted a three point function with LZ operators to a
computation involving DK operators. This has been computed in \DOT\KIT.
Note that we have assumed that negative number of screenings operators in the
\lio\ sector do not affect any of the contour deformation arguments. This can
be justified by first doing these operations and then analytically continuing
to negative screening.
In the \lio\ sector, the number of screenings  remains the same as we have
not moved out of the original Fock tower.

It is clear that the process would work out for the general case also, where
we have a set of three operators such that the total ghost number is zero.
Note that ghost number conservation for LZ operators translates to tower
number conservation for DK type operators.

The case of two-point functions is particularly simple . We can compute the
norm as in \IMM\ and as we have seen it is manifestly non-zero. In terms of
operators it is particularly easy to compute as the vertex operator charge of
the matter and Liouville sectors  precisely differ by a $2\al_0$ and a
$2\beta_0$ when we write them as DK type operators, thus requiring no
screening in either sector.
\newsec{Discussion}
We have made clear the relation between the LZ states
and the DK states and how correlators with one kind of operators
is  mapped into correlatiors of the other kind.
It is clear that the calculation of 3-point functions in the language of
DK operators really implements the non-decoupling of null states as
pointed out by Polyakov \POL. This in a calculation with only LZ operators
would be difficult to see. Further, the LZ operators involve Virasoro
secondaries and are therefore non-covariant in their construction. Here by the
explicit mapping from LZ states to DK states, we have provided a
covariantisation prescription implemented through the use of $Q_B$.
 However several questions remain to which we now turn. In a calculation
with DK states we can set up a class of 3-point functions with only positive
number of screening operators. This happens, for instance,in the $c_M=0$ case
for all
operators with $\beta <\beta_0$. This correlation function can be calculated
using the Coulomb gas formalism. However this cannot be mapped to a three-point
function with only LZ operators. In fact, the corresponding LZ states are all
of  negative ghost number and the three-point function of such objects would be
expected to be zero.It is not clear how ghost number conservation could be
 effectively violated in a theory with LZ operators alone. On the other hand,
a correlation function of LZ operators would if transformed to a correlation
function of DK operators, necessarily require negative screening in the
Liouville sector. Thus ghost number conservation may be intimately related to
negative screening in the \lio\ sector.

A similar set of descent equations  can clearly arise in the case of $W_N$
gravity.
It has been pointed out by Kalyana Rama\KAL\ that the Ising fermion
appears in $W_3$ gravity as a LZ type state of ghost number $+1$. We expect
that the same descent procedure, using the screening operators of $W_3$ and
$Q_B$, would relate the LZ state to a DK state at ghost number zero.
It would be interesting to apply our descent procedure in the case of
$W_N$ gravity coupled to matter.

The other interesting point is the fact that the mapping from LZ to DK states
relates massive states to massless states at specified momenta. A similar
feature has been noticed in the analysis of states in the black hole string
 theory by Distler and Nelson \DN. Indeedv, one may look for such features
in all cases where there is a non-trivial tachyon and dilaton background in
 critical strings.
If the OPE of such massless states generates a non-trivial symmetry algebra,
this symmetry algebra may in fact be understood as a symmetry of at least a
subset of the tower of massive states of the theory.

There is a curious analogy here with the work of Distler on $c=-2$ coupled to
\lio\JD. There the operators  in the $-1$ picture are simple vertex operators
 of DK type. However, the operators may be picture-changed to the $n-1$ picture
for the $\CO_n$ operator, in Distler's notation. The correlations can be
computed in the $-1$ picture with
appropriate  screening operators being added. In our setting, the DK type
states
are  all in a `fixed picture,' measured by the difference of the charges of
matter  and \lio\ vertex operators, which is, in fact, zero. The LZ states are
in different pictures, the difference in Liouville and matter charge
varying with ghost number and the level of the null state in the matter sector
that is dressed. However the operators have non-zero ghost number making the
analogy  somewhat imprecise.

\vskip.8truein{We would like to thank Swapna Mahapatra, Parameswaran Sankaran
and Ashoke Sen for useful discussions.}
\listrefs
\bye